\documentclass[a4paper,12pt,final]{article}

\usepackage[sort&compress]{natbib}
\setcitestyle{open={},close={},super}
\usepackage{hyperref}
\usepackage{amsmath}
\usepackage[T1]{fontenc}       
\usepackage{graphicx}
\usepackage{authblk}
\usepackage{bookmark}

\newcommand*\mycaption[1]{\caption{\small{#1}}}

\author[1]{Nara Rubiano da Silva}
\author[1]{Marcel M\"oller}
\author[1]{Armin Feist}
\author[2]{Henning Ulrichs}
\author[1]{Claus Ropers}
\author[1,3]{Sascha Sch\"afer\footnote{Contact author: sascha.schaefer@phys.uni-goettingen.de}}
\affil[1]{University of G\"ottingen, IV. Physical Institute, G\"ottingen, Germany}
\affil[2]{University of G\"ottingen, I. Physical Institute, G\"ottingen, Germany}
\affil[3]{Carl von Ossietzky Universit\"at Oldenburg, Institute of Physics, Oldenburg, Germany}

\date{}

\title{Nanoscale mapping of ultrafast magnetization dynamics with femtosecond Lorentz microscopy}

\begin{document}
	
\maketitle

\addcontentsline{toc}{section}{Abstract}
\begin{abstract}
Novel time-resolved imaging techniques for the investigation of ultrafast nano\-scale magnetization dynamics are indispensable for further developments in light-controlled magnetism. Here, we introduce femtosecond Lorentz microscopy, achieving a spatial resolution below \mbox{100 nm} and a temporal resolution of \mbox{700 fs}, which gives access to the transiently excited state of the spin system on femtosecond timescales and its subsequent relaxation dynamics.  We demonstrate the capabilities of this technique by spatio-temporally mapping the light-induced demagnetization of a single magnetic vortex structure and quantitatively extracting the evolution of the magnetization field after optical excitation. Tunable electron imaging conditions allow for an optimization of spatial resolution or field sensitivity, enabling future investigations of ultrafast internal dynamics of magnetic topological defects on 10-nanometer length scales.
\end{abstract}


\addcontentsline{toc}{section}{Main text}

Aiming for higher bit densities, manipulation of magnetization on the nanoscale is a key aspect for future information processing and storage applications. In recent years, significant achievements on controlling magnetic domains and topological defects on scales down to few hundreds of nanometers were obtained by various stimuli such as electrical current and light \cite{ruotolo2009, jiang2015, vanwaeyenberge2006, yu2010, torrejon2017, pollath2017, finazzi2013, eggebrecht2017, leguyader2012, liu2015}. Optical control of magnetization is particularly appealing due to the absence of an applied external field and the possibility for ultrafast switching speeds. The physical processes involved, such as direct spin-light interactions, optically-driven spin currents, and spin-flip scattering and the collapse of exchange splitting in a hot electron environment, remain active fields of study \cite{kimel2005, hansteen2005, koopmans2010, battiato2010, john2017}. For a further progress, it is essential to provide experimental tools to assess the optically-induced magnetization reordering processes at their intrinsic nanometer spatial and femtosecond temporal scales.

Time-resolved implementations of established experimental techniques for mapping magnetic structures at sub-micrometer dimensions have been already accomplished for magneto-optical Kerr effect microscopy \cite{mozooni2014}, photoemission electron microscopy \cite{choe2004}, scanning electron microscopy with polarization analysis \cite{fromter2016}, scanning transmission x-ray microscopy \cite{bisig2016, vanwaeyenberge2006}, small-angle x-ray scattering \cite{pfau2012}, and holography using x-rays \cite{buttner2015}. Novel imaging approaches using circularly  polarized high-harmonic radiation  \cite{kfir2017} may even provide access to magnetization dynamics on timescales of few femtoseconds.

Nanoscale mapping of static magnetic structures by Lorentz microscopy in a transmission electron microscope is a powerful method which routinely achieves resolutions down to few tens of nanometers \cite{degraef2000, raabe2000, jin2017}, with the possibility to correlate magnetization and structural information within the same instrument. Moreover, due to the direct effect of magnetic fields on the imaging electron wave, Lorentz image contrast allows for a quantitative reconstruction of the magnetization field \cite{degraef2000, volkov2002, eggebrecht2017}. 

Time-resolved Lorentz microscopy was previously demonstrated for investigations of field-assisted or laser-excited domain wall movement \cite{bostanjoglo1980, park2010, schliep2017}. Bostanjoglo and coworkers achieved nanosecond temporal resolution by electronically chopping the electron illumination beam \cite{bostanjoglo1980}. Obtaining shorter electron pulses became readily accessible with the advent of laser-triggered electron sources \cite{domer2003, browning2012, zewail2010}, extending Lorentz microscopy into the nanosecond \cite{park2010} and picosecond \cite{schliep2017} regime. However, the spatio-temporal resolution required to investigate ultrafast magnetic processes, such as the optical control of spin structures, necessitates advanced nanoscale photocathode approaches, providing femtosecond electron pulses with high spatial beam coherence \cite{feist2017}. 

Here, we introduce real-space nanoscale mapping of light-induced magnetization dynamics by ultrafast transmission electron microscopy (UTEM) with femtosecond temporal re\-solution. We quantitatively track the time-dependent magnetization field in a single vortex structure during laser-driven ultrafast demagnetization, reaching a 700-fs temporal and below-100-nm spatial resolution. These results demonstrate the capability of femtosecond Lorentz microscopy for the imaging of transient magnetization fields on timescales faster than the spin-lattice equilibration.

In our experiments, we study an isolated magnetic permalloy disc (\mbox{1 $\mu$m} diameter, \mbox{20 nm} thickness), prepared by electron-beam lithography on a silicon nitride membrane (50 nm thickness). The ground state texture of the disc consists of a magnetic vortex state, for which an in-plane oriented magnetization field $\vec{M}(\vec{r})$ of constant magnitude ${|\vec{M}(\vec{r})| = M_s}$ curls around the center of the structure \cite{hubert2000, shinjo2000, raabe2000}. In the vortex core, with a typical diameter on the 10-nm scale \cite{usov1993}, the magnetization turns to an out-of-plane direction.

\begin{figure}[h!]
	\centering 
	\includegraphics{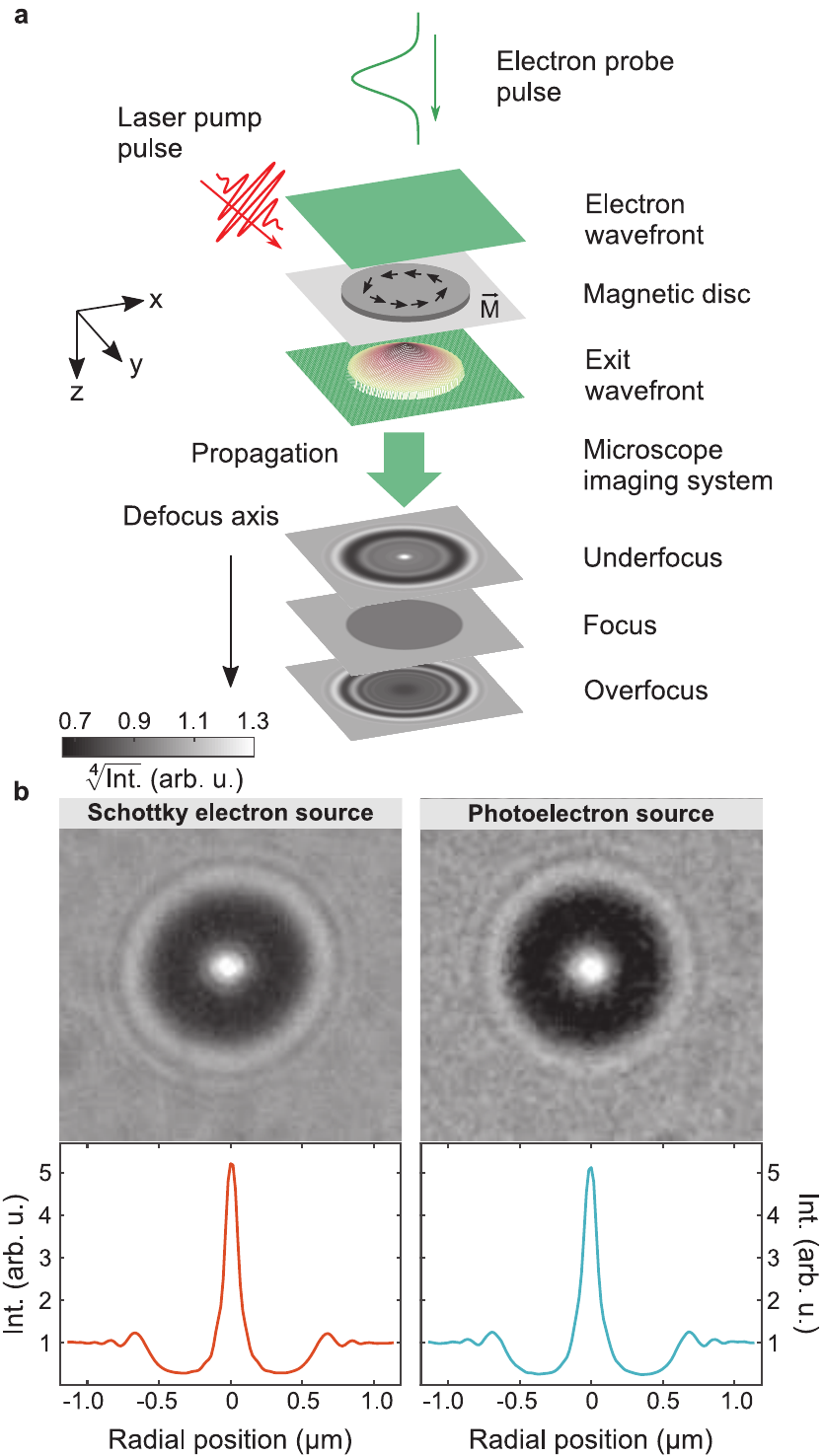}
	\mycaption{Femtosecond Lorentz microscopy. (a) Scheme of experimental setup and Lorentz image formation. (b) Lorentz micrographs of the permalloy disc recorded with fixed microscope conditions (\mbox{-9 mm} defocus) using Schottky emission and laser-triggered photoemission. Profiles in the lower panels show the azimuthally averaged intensity of the corresponding micrographs.}
	\label{fig:exp-setup}
\end{figure}

Transmission electron microscopy under out-of-focus imaging conditions, so-called Fresnel mode Lorentz microscopy \cite{degraef2003}, provides for an image contrast which is sensitive to the in-plane magnetization field (Fig.~\ref{fig:exp-setup}a). For the vortex sample, a conically-shaped Aharonov-Bohm phase shift $\phi$ \cite{aharonov1959} is imprinted onto an incident electron wavefront,

\begin{equation}
\phi = - \dfrac{e}{\hbar} \int{ \vec{A} \cdot \mathrm{d}\vec{s}},
\end{equation}

\noindent where $\vec{A}$ is the magnetic vector potential, $e$ the electron charge, $\hbar$ the reduced Planck constant, and the integral is computed along the electron beam trajectory. In defocused electron images, this spatial phase information is transferred to changes in electron image intensity, which, in general, can be employed to reconstruct the magnetization field by using phase retrieval approaches such as the transport-of-intensity equation \cite{degraef2000, volkov2002, eggebrecht2017}. For the magnetic vortex state considered here (Fig.~\ref{fig:exp-setup}b, upper left panel), the most prominent feature in the defocused electron micrograph is a bright or dark spot in the center of the magnetic disc. This maximum or minimum in electron intensity is caused by the lensing effect of the conical phase shift, which depends on the curling direction of the magnetic vortex and the sign of defocus. Furthermore, for a given defocus value, the image intensity in the disc center is a quantitative measure of the saturation magnetization $M_s$ (see below).


In order to perform time-resolved stroboscopic Lorentz microscopy, we developed a pulsed electron source based on nano-localized photoemission from a Schottky field emitter with a tunable pulse duration reaching \mbox{200 fs} \cite{feist2015, feist2017}. Depending on pulse charge and duration, this approach yields electron pulses  of high transverse coherence facilitating phase-contrast ima\-ging. The present experiments are  performed at a \mbox{500 kHz} repetition rate, and employ pulses with a duration of \mbox{700 fs}, containing up to \mbox{1 electron/pulse} at the sample. Figure \ref{fig:exp-setup}b (right panels) displays a Lorentz micrograph of the vortex structure collected with a photoelectron beam. The visibility of Fresnel fringes at the edge of the disc is largely retained, demonstrating comparable spatial coherence of the photoelectron source to a thermal Schottky emitter (Fig.~ \ref{fig:exp-setup}b, left panels). For the pulsed photoelectron illumination utilized in the time-resolved experiments, a transverse coherence length of \mbox{76 nm} is extracted (see Supplement).

\begin{figure}
	\centering\includegraphics{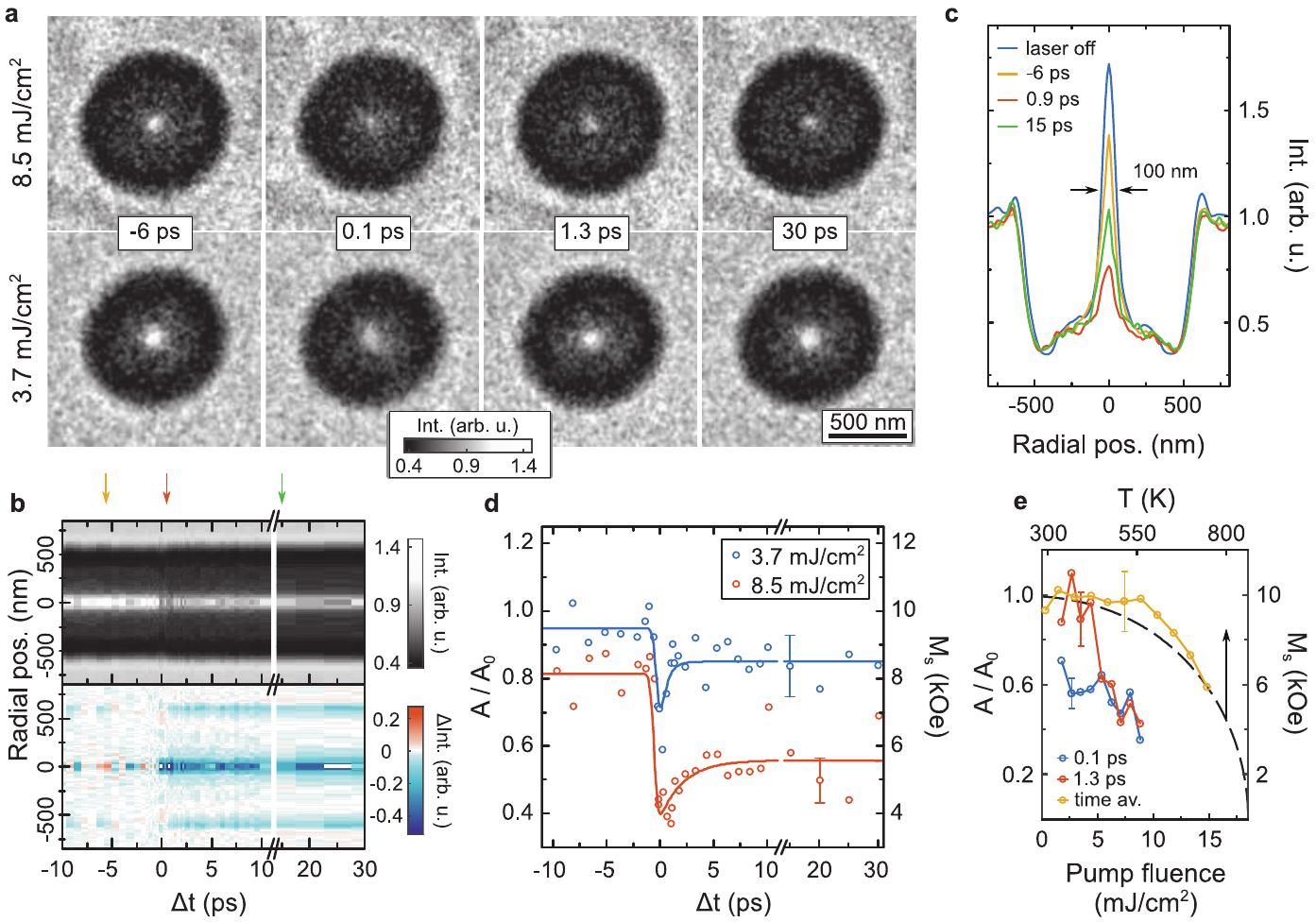}
	\mycaption{Delay-time-dependent Lorentz contrast of a single, optically excited magnetic disc. (a) Time-resolved Lorentz micrographs for pump fluences of 8.5 (top row) and \mbox{3.7 mJ/cm$^2$} (bottom row). (b,c) Temporal evolution of azimuthally averaged intensity profiles of the micrographs for a pump fluence of \mbox{8.5 mJ/cm$^2$} (top row). In the bottom row, the mean profile before the arrival of the pump pulse is subtracted. Profiles for the marked delay times are shown in (c), along with a profile of the unpumped disc. (d) Temporal evolution of the image amplitude $A$ at the disc center (normalized by the amplitude $A_0$ for the unpumped structure). Solid lines: Exponential recovery model. (e) The amplitude $A$ as a function of pump fluence from micrographs obtained using a continuous electron beam and electron pulses at specific time delays. Black dashed curve: Weiss molecular field theory adapted to the data.}
	\label{fig:lorentzUTEM}
\end{figure}

We optically excite the magnetic vortex structure with femtosecond laser pulses (\mbox {50 fs} duration, \mbox{800 nm} center wavelength, about \mbox{50 $\mu$m} full-width-at-half-maximum (FWHM) spot diameter, p-polarized at a 55$^{\circ}$ angle of incidence), and probe the transient magnetic structure with ultrashort electron pulses as a function of delay time $\Delta t$ between pump and probe pulses. Figure \ref{fig:lorentzUTEM}a shows a set of Lorentz micrographs (6 minutes exposure time each, \mbox{-4.5 mm} defocus) for varying delay times and two optical excitation fluences. Directly after excitation (\mbox{$\Delta t = 0.1$ ps}), we observe a strong decrease in the intensity of the bright central spot, followed by a partial recovery on a picosecond timescale. Selected image intensity profiles (azimuthally averaged around the vortex core) are shown in Fig.~\ref{fig:lorentzUTEM}c.


Considering a larger number of delay times, Fig.~\ref{fig:lorentzUTEM}b displays the time evolution of the averaged radial intensity profiles of the Lorentz micrographs (upper panel) and the corresponding image intensity changes (lower panel). Adapting Lorentzian line profiles with amplitude $A$ to the intensity near the disc center, we extract the delay-dependent normalized amplitude ${A(\Delta t)/A_0}$ ($A_0$: mean amplitude without optical excitation). For an analysis of the delay-dependent traces (Fig.~\ref{fig:lorentzUTEM}d), we adopt an exponential recovery model (Fig.~\ref{fig:lorentzUTEM}d, solid lines) given by:

\begin{equation}
A(\Delta t) = C - H(\Delta t) [a + b(1 - \rm{e}^{-\Delta t/t_{rec}}) ],
\label{eq:model}
\end{equation}

\noindent where $H(\Delta t)$ is the Heaviside step function, $C$ the amplitude for negative delay times, $a$ the amplitude of the initial drop, and $b$ the amplitude of the partial magnetization recovery at later delay times. The finite electron pulse duration is included by convolution of the expression in Eq.~\ref{eq:model} with a corresponding Gaussian envelope. The time traces are well described by recovery time constants of about $t_{rec}=0.6$ and 2.2 ps for the low and high fluence case, respectively.

In a simple three-temperature-model (3TM) \cite{beaurepaire1996, koopmans2010}, which considers individually thermalized electron, spin and lattice subsystems, optical excitation results in a ultrafast drop of magnetization due to rapid electron-spin coupling on a timescale of a few 100 fs. The subsequent coupling with the lattice bath leads to a partial recovery of the sample magnetization. The corresponding timescale $t_{rec}$ on the order of 1 ps is resolved in our experiments and is in agreement with previous ultrafast optical spectroscopy experiments  \cite{mathias2012, walowski2008}.

\begin{figure}
	\centering\includegraphics{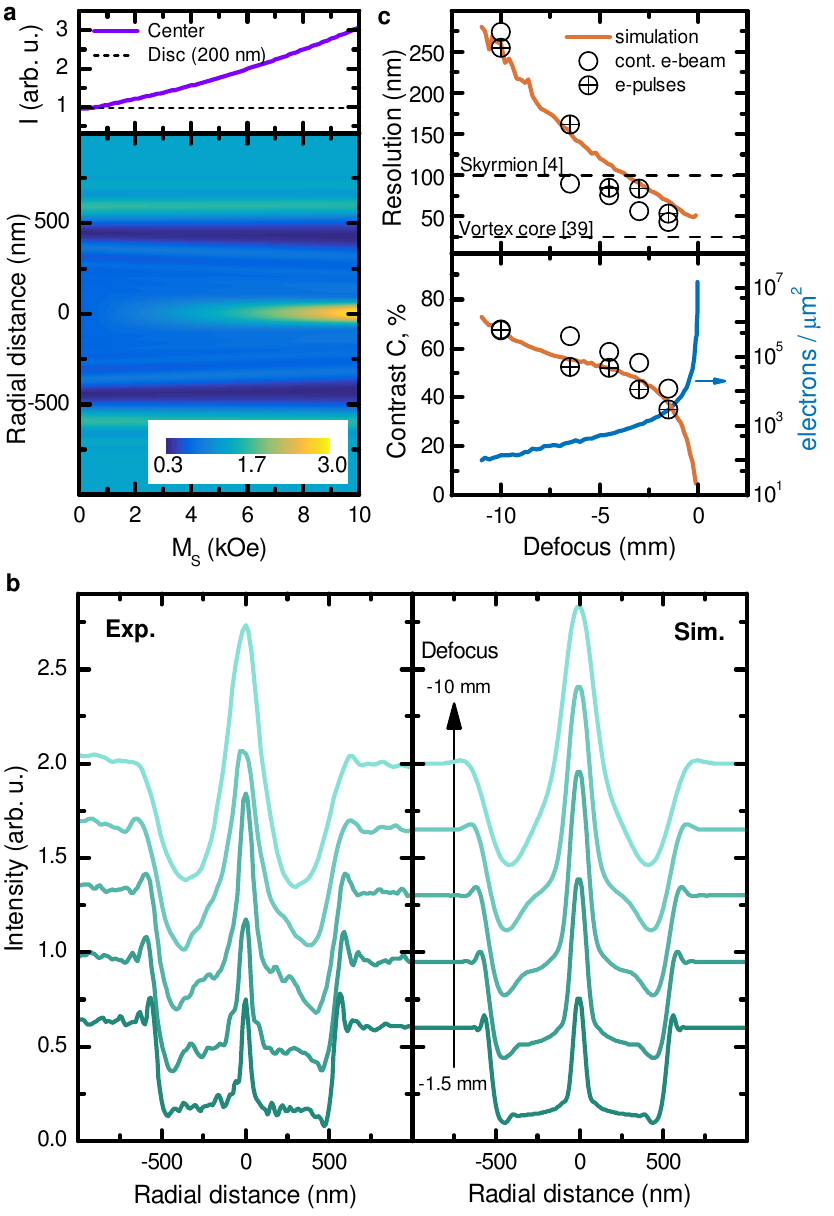}
	\mycaption{Spatial resolution and quantitative image analysis in femtosecond Lorentz microscopy. (a) Radial profiles of simulated Lorentz images (lower panel) demonstrate an approximate proportionality between the image intensity at the disc center and saturation magnetization (upper panel). (b) Radial profiles of experimental micrographs (left panel) and simulated images (right panel) at defoci of -10, -6.5, -4.5, -3.0 and \mbox{-1.5 mm}. Micrographs were obtained using the same illumination conditions and electron pulse durations as for the time-resolved experiments.  (c)  Resolution (upper panel, full-width-at-half-maximum $w$ of the central peak) and contrast (lower panel, left axis, comparing the intensities $I_C$ at the center and $I_{disc}$ at 200 nm from the center) extracted from the radial profiles for different defocus values. From the contrast $C$ in the simulated images, we estimate the number of electrons required for resolving the bright spot of area $\pi w^2/4$ just above the shot noise level as $1/C^2$ (lower panel, right axis).}
	\label{fig:simLim}
\end{figure}

A unique feature of ultrafast Lorentz microscopy is the direct connection of image intensity changes to the transient magnetization distribution. For a quantitative analysis, we perform electron image simulations (see Supplement) for a vortex state with varying saturation magnetization $M_s$ (with fixed ${\vec{M}(\vec{r}) / M_s}$)\footnote{Generally, the magnetic configuration is expected to respond to an increased spin temperature in a complex manner due to changes, for example, in the average dipolar and exchange interactions. For the present case, we performed micromagnetic simulations which suggest that only changes in the saturation magnetization need to be considered on ultrashort time scales (see also Supplement).}. The dependence of the radial image profiles on the saturation magnetization $M_s$ (Fig.~\ref{fig:simLim}a) demonstrates that, for the chosen imaging conditions, the amplitude $A$ of the central image spot is approximately proportional to $M_s$. 

Using a linear relation between $A$ and $M_s$, we can now extract the delay-dependent magnetization ${M_s (\Delta t)}$ from the fluence-dependent transient amplitude changes in Figs.~\ref{fig:lorentzUTEM} d,e. The time-averaged saturation magnetization (Fig.~\ref{fig:lorentzUTEM}e, yellow curve) follows a behavior which can be described within the Weiss molecular field theory \cite{kittel2004} (broken black curve) and originates from an average sample heating under laser illumination. From the time-resolved Lorentz micrographs at $\Delta t=0.1$~ps, we extract a drop in saturation magnetization to about $3.7-7.0 \pm 0.7$~kOe in the considered fluence range. For example, for a fluence of \mbox{8.5 mJ/cm$^2$} (Fig.~\ref{fig:lorentzUTEM}d, red curve), a minimum magnetization of  $3.7 \pm 0.7$~kOe is found for the hot spin system at $\Delta t=0.9$~ps, recovering to $5.6 \pm 0.7$~kOe at longer delay times (corresponding temperature of about \mbox{753 K}).


These results demonstrate the capabilities of ultrafast Lorentz microscopy for the nanoscale mapping and quantitative characterization of magnetization dynamics, and suggest its application for high-resolution real-space investigations of ultrafast magnetic phenomena. To further gauge the achievable spatial resolution, we consider simulated Lorentz images of the vortex core region at different defoci and compare them to experimental Lorentz micrographs. In Fig.~\ref{fig:simLim}b, we plot the simulated Lorentz image profiles for varying defocus, which are in convincing agreement with the experimental data. The vortex in the present geometry is estimated to have a core diameter of about \mbox{26 nm} \cite{usov1993}. At the -4.5-mm defocus adopted in the time-resolved experiments, the central image spot is broadened to about \mbox{100 nm} (FWHM), and the detailed structure of the magnetization field close to the core is largely unresolved. The resolution is considerably enhanced for smaller defoci, but with reduced image contrast. Figure~\ref{fig:simLim}c shows the resolution and contrast of the vortex core region when imaging with electron pulses (circle plus symbols) and with Schottky-emitted electrons (open circles). The comparable resolution in both cases demonstrates that transient magnetic structures can be imaged with similar quality to conventional Lorentz microscopy. However, at small defoci a higher electron dose is required (Fig.~\ref{fig:simLim}c, lower panel) to resolve changes in the low contrast images. For the current electron pulse parameters and at a defocus of -1.5 mm, we could achieve a resolution of \mbox{$55 \pm 2$ nm} within a reasonable 5 minutes exposure time. With such a spatial resolution in femtosecond Lorentz microscopy now available, a broad range of ultrafast magnetic processes are accessible, including high-frequency internal dynamics of domain walls and  magnetic Skyrmions \cite{montoya2017}.

In conclusion, we demonstrated the mapping of ultrafast demagnetization dynamics in a single nanostructure with sub-100-nm-spatial and 700-fs-temporal resolution by femtosecond Lorentz microscopy. Transient magnetization changes are extracted in a quantitative manner, rendering femtosecond Lorentz microscopy a unique tool to study non-equilibrium spin nano structures.

\addcontentsline{toc}{section}{Acknowledgement}
\section*{Acknowledgement}
We gratefully acknowledge funding by the Deutsche Forschungsgemeinschaft (DFG-SPP-1840 \lq{Quantum Dynamics in Tailored Intense Fields}\rq, and DFG-SFB-1073 \lq{Atomic Scale Control of Energy Conversion}\rq, projects A05 and A06), support by the Lower Saxony Ministry of Science and Culture and funding of the instrumentation by the DFG and VolkswagenStiftung. N.R.S. would like to acknowledge the support by the Conselho Nacional de Desenvolvimento Cient\'{i}fico e Tecnol\'{o}gico (Science Without Borders Program, Governo Dilma Rousseff, Brazil).





\addcontentsline{toc}{section}{Authors contributions}
\section*{Authors contributions}

N.R.S. conducted the time-resolved Lorentz microscopy experiments with contributions from A.F., analyzed the data and implemented Lorentz image simulations with contributions from S.S. and C.R.. M.M. prepared the magnetic disc sample. H.U. carried out micromagnetic modelling of the sample system. N.R.S. and S.S. wrote the manuscript with contributions from all authors. C.R. and S.S. conceived and directed the study. All authors discussed the results and the interpretation.


\addcontentsline{toc}{section}{Supporting Information}
\section*{Supporting Information}

\addcontentsline{toc}{subsection}{SI1: Movies of ultrafast demagnetization in a single vortex structure}
\subsection*{SI1: Movies of ultrafast demagnetization in a single vortex structure}

Lorentz microscopy movies of the light-induced demagnetization of a single vortex structure were generated from the stack of time-resolved micrographs analysed in the main text (cf. Fig.~2 in the main text, and SI 3). Movie M1 displays the delay-dependent Lorentz micrographs for a pump fluence of  8.5~mJ/cm$^2$ (cf. Fig.~2b in the main text, upper panel). For comparison, M2 shows the corresponding image contrast changes for a fluence of 3.7~mJ/cm$^2$.


\addcontentsline{toc}{subsection}{SI2: Experimental details}
\subsection*{SI2: Experimental details}

The experiments were realized with the Ultrafast Transmission Electron Microscope (UTEM) which we recently developed at the University of G\"ottingen \cite{feist2017}. Ultrafast temporal resolution is achieved in a stroboscopic manner by a laser-pump/electron-probe scheme. The instrument consists of an amplified femtosecond laser system (800-nm center wavelength, 50-fs pulse duration, operated here at a 500-kHz repetition rate) with two optical beam paths coupled to a field emission TEM (JEOL JEM-2100F). In the first beam path, optical pulses are frequency-doubled and focused onto a Schottky-type photoemitter for generating femtosecond electron pulses. In the second beam path, 800-nm laser pulses are focussed on the sample (at 55$^{\circ}$ incidence angle relative to the sample plane, p-polarized, 50-$\mu$m focal spot size) and trigger ultrafast demagnetization. A variable attenuator allows for a tuning of the pump pulse power, and the timing between the electron probe and laser pump pulses is controlled by a mechanical delay stage.

For a Schottky-type emitter, linear photoemission can be localized to the nanoscopic front facet of the emitter tip, yielding electron pulses with high transverse coherence \cite{feist2017}. The electron pulse charge and duration can be tuned by the intensity of the photoemission laser pulse and the electric extraction field at the tip apex. Here, we used an extraction field of about \mbox{1 V/nm}, and a \mbox{2.3-nJ} optical pulse energy (estimated spot diameter on the tip of \mbox{20 $\mu$m}) to generate 700-fs electron pulses at an acceleration voltage of 120 kV. At the sample plane, the electron pulses contained up to one electron, and the illumination was spread to a 25-$\mu$m-diameter region with the magnetic vortex in the center.


\addcontentsline{toc}{subsection}{SI 3: Image acquisition and data analysis}
\subsection*{SI 3: Image acquisition and data analysis}

For recording Lorentz micrographs, we employ an image mode in which the main objective lens is turned off (\lq{Low Magnification\rq} mode, 6000-fold nominal magnification), so that the sample is in a low-magnetic-field environment (residual field about \mbox{19 mT}, see Ref.~\citenum{kohn2012}).  Image defocus was adjusted by changing the excitation of a subsequent magnetic lens (\lq{objective mini lens}\rq). Each image was integrated for 1 minute (sum of 6 individual exposures, \mbox{10 s} each, total dose of about 23000 \mbox{electrons/$\mu$m$^2$}, \mbox{13-nm} effective pixel size). The time delay scans consisted of 52 steps covering the range from \mbox{-10 ps} to \mbox{30 ps}, and were repeated 6 times. Delay times were scanned in a random sequence. To account for variations in electron beam current, each image was normalized by the total image intensity. Slow image drifts are corrected for by applying an optimized translation vector which minimizes the mean square change of pixel intensities (integrated over the whole frame, relative to a reference frame). For the time-resolved data shown in the main text, we evaluate the median intensities of each image pixel across the 6 repetitions per delay time. Finally, we applied to each frame a Gaussian spatial filter with a width of 0.7 pixel to further reduce image noise.

For the radial intensity profiles $I(r)$ around the vortex center, we computed the total intensity within rings of radius $r$ (1-pixel width), normalized to the number of pixels within each ring. For the evaluation of the image intensity maximum in the center of the vortex structure, the radial intensity profiles (for \mbox{$r < 250$ nm}) were fitted using a Lorentzian model ${I(r) = I_0 + \frac{2a}{\pi} \frac{w}{4 (r-r_0)^2 + w^2}}$ centered at ${r = r_0}$, with offset $I_0$,  area $a$ and full-width-at-half-maximum $w$. For quantitative analysis of the transient Lorentz contrast  shown in Figs.~2d,e in the main text, we considered the amplitude ${A = 2a/(\pi w)}$. The full-width-at-half-maximum was used to gauge the resolution of the magnetic contrast. The same analysis procedure was applied to the experimental and simulated images of the defocus series (Figs.~3b-c in the main text).


\addcontentsline{toc}{subsection}{SI 4: Lorentz image simulation}
\subsection*{SI 4: Lorentz image simulation}

For calibration of the Lorentz contrast and to estimate the spatial resolution limits, we simulated Lorentz images of the magnetic vortex structure at different values of the saturation magnetization and at different defoci. To this end, we consider the propagation of an ensemble of electron waves in the electron-optical imaging system of the microscope \cite{degraef2003,lichte2008}, as explained below.

The illuminating electron wave function $\psi_0$ is spatially modified in amplitude and phase by electron-sample interactions. Here, we consider $\psi_0$ to be a plane wave (parallel illumination). In Lorentz microscopy,  the most relevant interactions (i.e. with the electromagnetic field inside the sample) induce a phase shift $\phi$ on $\psi_0$, while other scattering events can be included by a spatially dependent amplitude modulation $A(\vec{r})$. In the projection approximation, the exit wave function $\psi$ below the sample plane is given by

\begin{equation}
\psi (\vec{r}) = |A(\vec{r}) |\ \mathrm{e}^{\mathrm{i} \phi(\vec{r})} \psi_0,
\end{equation}

\noindent where $\vec{r}$ is a vector in the plane perpendicular to the microscope optical axis (xy-plane, see Fig.~1 in main text for the coordinate system). The amplitude modulation is extracted from in-focus images. The phase shift $\phi$ is related to the magnetic structure by the Aharonov-Bohm equation \cite{aharonov1959}

\begin{equation}
\phi = \dfrac{e}{\hbar} \left ( \dfrac{1}{v} \int{V(\vec{r}, z) ds} - \int{ \vec{A}(\vec{r},z) \cdot \mathrm{d}\vec{s}} \right ),
\end{equation}

\noindent where $v$ is the speed of the imaging electrons and $V$ and $\vec{A}$ are the electrostatic and magnetic potentials inside the sample. The integrals are computed along the electron beam direction~$z$.

The electrostatic component of the phase shift within the disc is calculated by taking into account a thickness \mbox{$t = 20$ nm} for the permalloy film and a mean inner potential \mbox{$V = 36$ V}. For the vortex structure, the magnetic component of the phase shift can be closely approximated to exhibit radially-symmetric conical shape, which for ${r < R}$ is given by \cite{petford-long2002}

\begin{equation}
\phi_m  (r) =\pm \dfrac{\pi B_0 t}{\phi_0} (R- r),
\end{equation}

\noindent where $R$ is the disc radius, $\phi_0$ the magnetic flux quantum (\mbox{$\phi_0 = 2070$ T/nm$^2$}), and $B_0$ the in-plane magnetic flux density. For ${r > R}$, $\phi_m$ is zero. The positive (negative) sign corresponds to a counter-clockwise (clockwise) curling of the sample magnetization (viewed along the $+z$ direction).  For the simulation, we employed \mbox{$B_0 t = 12.24$ nm$\cdot$T}, reproducing the experimental Lorentz contrast.

For a known exit wave function $\psi (\vec{r})$, the image $I (\vec{r})$ produced by the electron optical system of the Lorentz microscope can be calculated via

\begin{equation}
I (\vec{r}) = |\mathrm{F}^{-1}\  [\mathrm{F}[\psi (\vec{r})]\ T(\vec{q}) ] | ^2,
\end{equation}

\noindent where $T$ is the transfer function \cite{degraef2003}, and $\vec{q}$ the wave vector (reciprocal of the spatial vector $\vec{r}$). For a coherent illumination, the transfer function including the lowest-order coherent aberrations of an astigmatism-corrected microscope is given by \cite{degraef2003, alexander1997}

\begin{equation}
T (\vec{q}) = \mathrm{e}^{-\mathrm{i} \chi (\vec{q}) } = \mathrm{exp}\left[ -\dfrac{\mathrm{i}2\pi}{\lambda}\  \left(\dfrac{\Delta f}{2}(\lambda q)^2 + \dfrac{C_s}{4}(\lambda q)^4 \right) \right],
\label{eq:ctf1}
\end{equation}

\noindent where $C_s$ is the spherical aberration constant of the objective (mini-)lens, $\Delta f$ the defocus (positive for overfocus), and $\lambda$ the electron wavelength.

The limited degree of temporal and spatial coherence of electron sources gives rise to incoherent aberrations, which affect image contrast. Firstly, the spread of electron wavenumbers $k = 1/\lambda$ (temporal coherence) results in varying focal lengths for different electron energies. The resulting electron image is an incoherent summation of image intensities (as given by Eq.~4) considering a distribution of electron wavenumbers. We include this contribution by multiplying the transfer function with a damping term (temporal coherence envelope) \cite{lichte2008}

\begin{equation}
E^{tc} (\vec{q}) = \mathrm{exp} \left[ - \dfrac{\pi^2}{2} \left( \lambda C_c\ \sqrt{\dfrac{\sigma^2_E}{(e U^*_a)^2}} \right)^2 q^4 \right],
\end{equation}

\noindent where $C_c$ is the chromatic aberration constant of the imaging lens and $\sigma^2_E$ is the standard deviation of the electron energy distribution. The relativistically modified acceleration potential is given by ${U^*_a = U_a \left( 1 + \frac{e U_a}{2 m_o c^2} \right)}$, in which $U_a$ is the experimental acceleration voltage, $m_0$ the electron rest mass and $c$ the speed of light.

Imaging utilizing illumination conditions with partial spatial coherence can be described in a similar manner. Specifically, waves with different incidence angles $\vec{\alpha}_{il}$ contribute incoherently to the final image $I (\vec{r})$, which is given by the sum of the images resulting from illumination with a single $\vec{\alpha}_{il}$, weighted by the probability distribution $i(k \vec{\alpha}_{il})$ of illumination directions \cite{lichte2008}:

\begin{equation}
I (\vec{r}) = \int_{source} i(k \vec{\alpha}_{il}) I_{k \vec{\alpha}_{il}}\ \mathrm{d}(k \vec{\alpha}_{il}).
\label{eq:incsum}
\end{equation}

For small tilt angles, the transfer function for tilted illumination can be approximated  using $\chi (\vec{q} + k \vec{\alpha}_{il}) = \chi (\vec{q} )\ +\ k \vec{\alpha}_{il} \nabla{\chi}(\vec{q} )$. Using this Taylor expansion and a rotationally symmetric Gaussian distribution of incidence angles $i(k \vec{\alpha}_{il})$, the image intensity can be calculated by modifying the transfer function by an additional damping term (spatial coherence envelope) \cite{lichte2008}:

\begin{equation}
E^{sc} (\vec{q}) = \mathrm{exp} \left[ -\pi^2 \dfrac{\theta_c^2}{\lambda^2 \mathrm{ln}(2)} \left( C_s (\lambda q)^3 + \Delta f (\lambda q)^2 \right) \right],
\end{equation}

\noindent where $\theta_c$ is the opening half-angle of the illumination. Within this approximation, the complete transfer function is:

\begin{equation}
T (\vec{q}) = E^{tc} (\vec{q}) E^{sc}(\vec{q}) \mathrm{e}^{-i \chi (\vec{q}) }.
\label{eq:ctf2}
\end{equation}

\begin{figure}[b!]
\begin{center}
	\includegraphics[width=\textwidth]{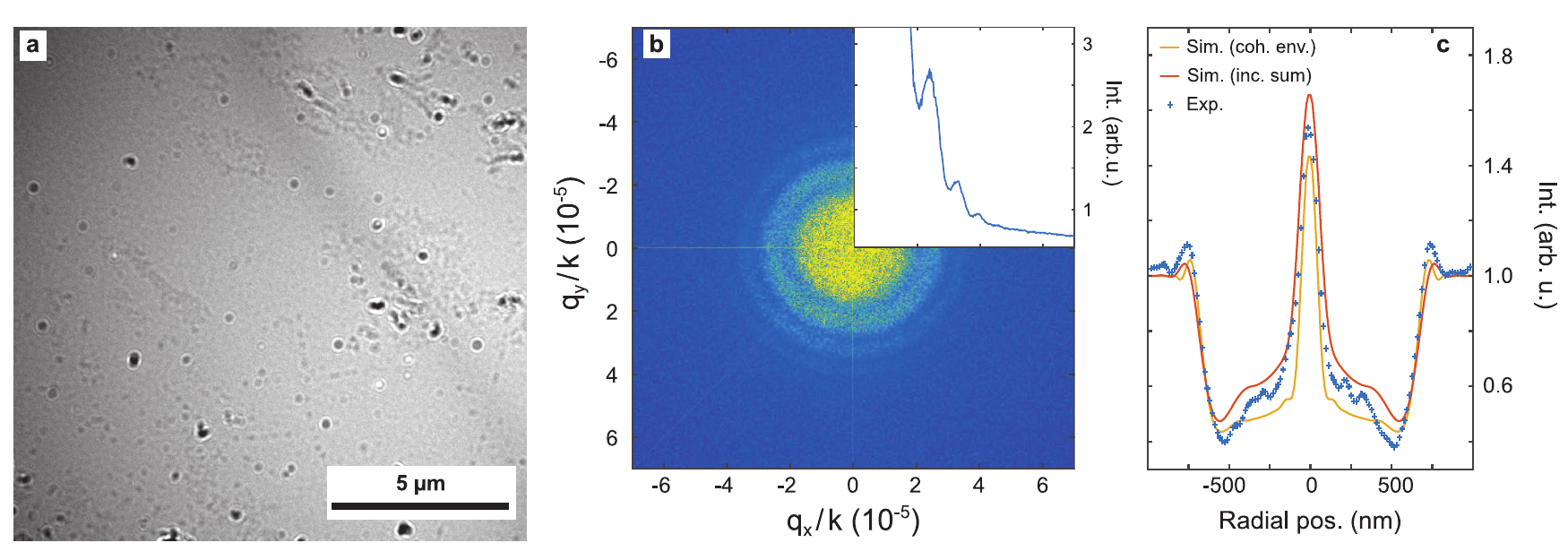}
	\caption{Imaging parameters and Lorentz image simulation. (a) Experimental Lorentz micrograph of the silicon nitride substrate recorded at a position far from the magnetic structure and (b) its respective diffractogram, which was used for calibrating defocus and spread of incidence angle. The micrograph was obtained using Schottky electrons at the same imaging conditions as used in the time-resolved experiments. The position and visibility of the Thon rings were obtained from the azimuthally averaged diffractogram (inset). (c) Radial profiles of simulated and experimental Lorentz micrographs of the magnetic vortex structure. Treating the spread of incidence angles by numerical  averaging (red curve) results in better agreement with the experimental data (symbols), compared to an approximative treatment using a spatial coherence envelope approach (yellow curve).}
	\label{supplfig:sim}
\end{center}
\end{figure}

For the large defocus values employed here, $\chi (\vec{q})$ exhibits rapid oscillations in $\vec{q}$, so that the lowest order Taylor expansion of Eq.~\ref{eq:ctf1} is only of limited validity. In a more precise approach, we numerically integrate Eq.~\ref{eq:incsum} to properly account for the illumination with partial spatial coherence. Specifically, we randomly sampled 2000 values for the incidence angle $\vec{\alpha}_{il}$ according to a gaussian distribution with a $2 \theta_c$ spread for the numerical integration of Eq.~\ref{eq:incsum}. A comparison of both image simulation approaches and an experimental Lorentz profile is shown in Fig.~\ref{supplfig:sim}c.

In the simulation, we adopt aberration constants of \mbox{$C_s = 4.61$ m} and \mbox{$C_c = 8.7$ cm} (reported for 200 keV electron energy \cite{kohn2012}), and an experimentally measured electron energy width of electron pulses of about \mbox{1.7 eV}. The defocus ${\Delta f}$ and the spread of incidence angles $\theta_c$ were obtained from the position and visibility of Thon rings (Fig.~\ref{supplfig:sim}b) in the Fourier transform of defocused micrographs of an amorphous silicon nitride substrate (Fig.~\ref{supplfig:sim}a). Focus-dependent calibration of the magnification was performed using the spacing between fabricated holes in the membrane. For the experimental conditions utilized in the time-resolved measurements, we found a defocus value of \mbox{-4.5 mm} and a spread of incidence angles \mbox{$\theta_c =7 \mu$rad} (corresponding to a transverse coherence length of \mbox{$\frac{\hbar}{m_0 c \beta \gamma} \frac{1}{\theta_c} = 76$ nm}, in which $\beta = v/c$ and $\gamma$ is the Lorentz factor \cite{vanoudheusden2007}). For the Lorentz micrographs shown in Figs.~3b and 3c in the main text, defocus values of \mbox{-10 mm}, \mbox{-6.5 mm}, \mbox{-3.0 mm} and \mbox{-1.5 mm} were found.

\addcontentsline{toc}{subsection}{SI 5: Micromagnetic simulation of the out-of-equili\-brium vortex structure}
\subsection*{SI 5: Micromagnetic simulation of the out-of-equili\-brium vortex structure}

\begin{figure}[h!]
\begin{center}
	\includegraphics[width=\textwidth]{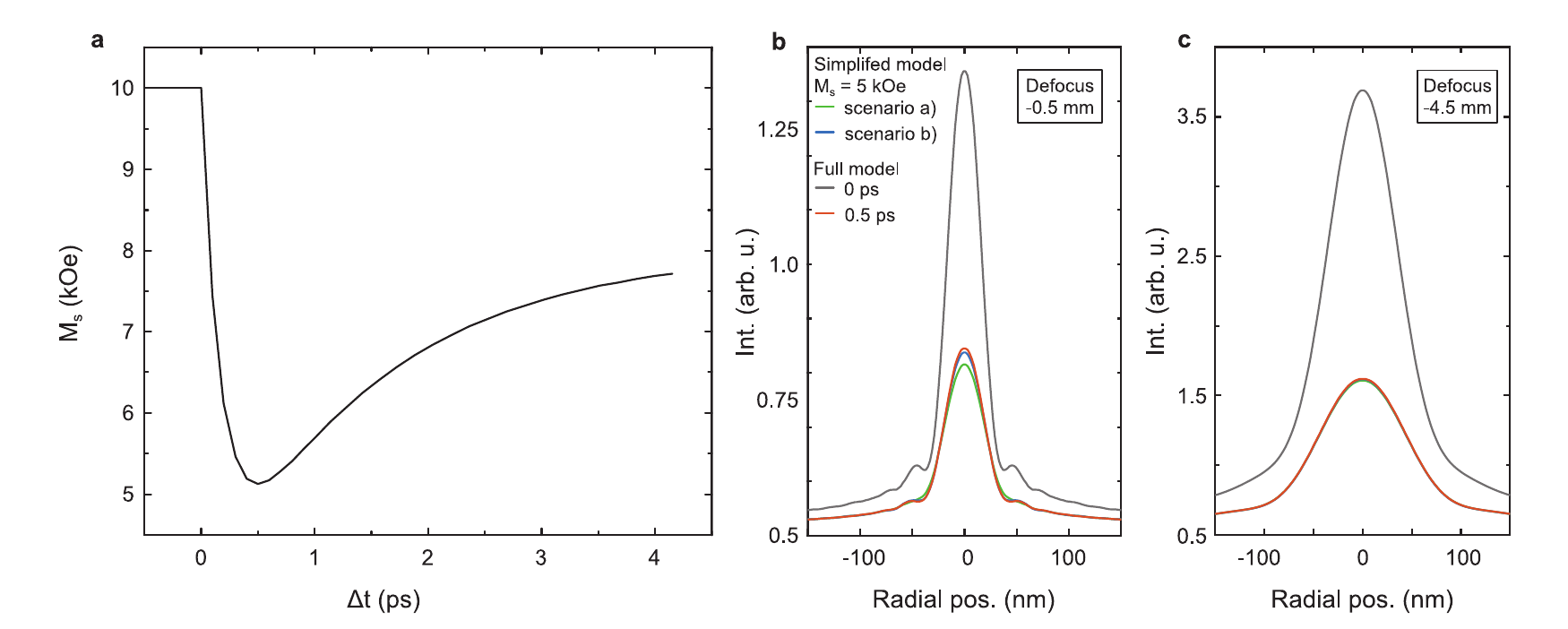}
	\caption{Micromagnetic simulation of the optically-excited vortex structure. (a) Ultrafast response of the saturation magnetization to the optical excitation, used to model the magnetization of the vortex structure on a picosecond time scale. (b-c) Radial profiles of simulated Lorentz micrographs from the micromagnetically calculated magnetization distributions. The micrographs were simulated using \mbox{7 $\mu$rad} spread of incidence angles (cf. Section SI 4) and a defocus of \mbox{(b) -0.5 mm} and \mbox{(c) -4.5 mm}, respectively.}
	\label{supplfig:micromsim}
\end{center}
\end{figure}

Time-dependent micromagnetic simulations were performed with the software package mumax3 \cite{vansteenkiste2014}. Similar to the experiment, ultrafast quenching of the magnetization was modelled in a 30-nm-thick, 1-$\mu$m-diameter permalloy disc. For the temporal behavior of the saturation magnetization $M_s = M_s(\Delta t)$ we apply a three-temperature model \cite{walowski2008} (Fig.~\ref{supplfig:micromsim}a), assuming a time-independent exchange constant and a homogenous excitation of the permalloy disc. 

For a hot spin system, the spatial structure of the magnetization field, such as the vortex core diameter, is expected to adapt due to changes, for example, in the dipole interaction. Generally, changes in the spatial structure will depend on the time scales involved.  In comparison to the time-dependent micromagnetic simulation, we consider two limiting scenarios: a) the magnetization field $\vec{M} (\vec{r})$ adapts adiabatically to the minimum energy configuration for the given saturation magnetization, and b) the local magnetization structure stays constant except for a homogeneous change of the length of the magnetization vectors. 

The simulated Lorentz micrographs (-0.5 mm defocus) for these cases are shown in Fig.~\ref{supplfig:micromsim}b. For a delay time of 0.5 ps, a constant magnetization structure (scenario b) more accurately describes the evolving field. On sub-picosecond timescales, spin waves, which mediate changes in the spatial magnetization structure, have only propagated over short distances. Only at later times t > 5 ps (not shown), a reorientation of the magnetization vectors becomes observable in the simulation. We note that for larger defocus values, information on high spatial frequencies is lost in the Lorentz images, and the different demagnetization scenarios are not distinguishable, as shown in Fig.~\ref{supplfig:micromsim}c.


\addcontentsline{toc}{subsection}{SI 6: Deriving the sample magnetization from Lorentz image contrast}
\subsection*{SI 6: Deriving the sample magnetization from Lorentz image contrast}

The Weiss molecular field theory describes the temperature dependence of the saturation magnetization as ${m = \mathrm{tanh}(m/t)}$ \cite{kittel2004}, where $m=M_s(T)/M_s(0)$ is the reduced magnetization and $t=T/T_c$ the reduced temperature (${m = 1}$ at 0 K and ${m = 0}$ at the Curie temperature $T_C = 850$~K \cite{wakelin1953}). With the bright spot intensity A in the Lorentz micrographs being proportional to $M_s$, we extract for a fluence of \mbox{8.5 mJ/cm$^2$}, a base temperature (i.e. for negative delay times) of \mbox{561 K} and an equilibrated temperature of \mbox{753 K} for longer delay times.

For comparison, we calculate the optically induced temperature rise of the permalloy disc from the incident pump fluence and the known material constants (for Ni, approximate composition of permalloy: 80\% Ni and 20\% Fe). For the laser illumination conditions and not considering near-field effects, we estimate an absorbance of the 20-nm-thick magnetic sample of about 25\%. With a heat capacity \mbox{$c_p = 0.445$ J/g.K} and a density \mbox{$\rho = 8.90$ g/cm$^3$} \cite{lide2002}, we find a temperature increase of 268 K, in reasonable agreement with the temperature estimate from the Lorentz contrast given above.


\addcontentsline{toc}{subsection}{SI 7: Electron-pulse temporal duration and its influence on Lorentz image contrast}
\subsection*{SI 7: Electron-pulse temporal duration and its influence on Lorentz image contrast}

\begin{figure}[h!]
\begin{center}
	\includegraphics[width=\textwidth]{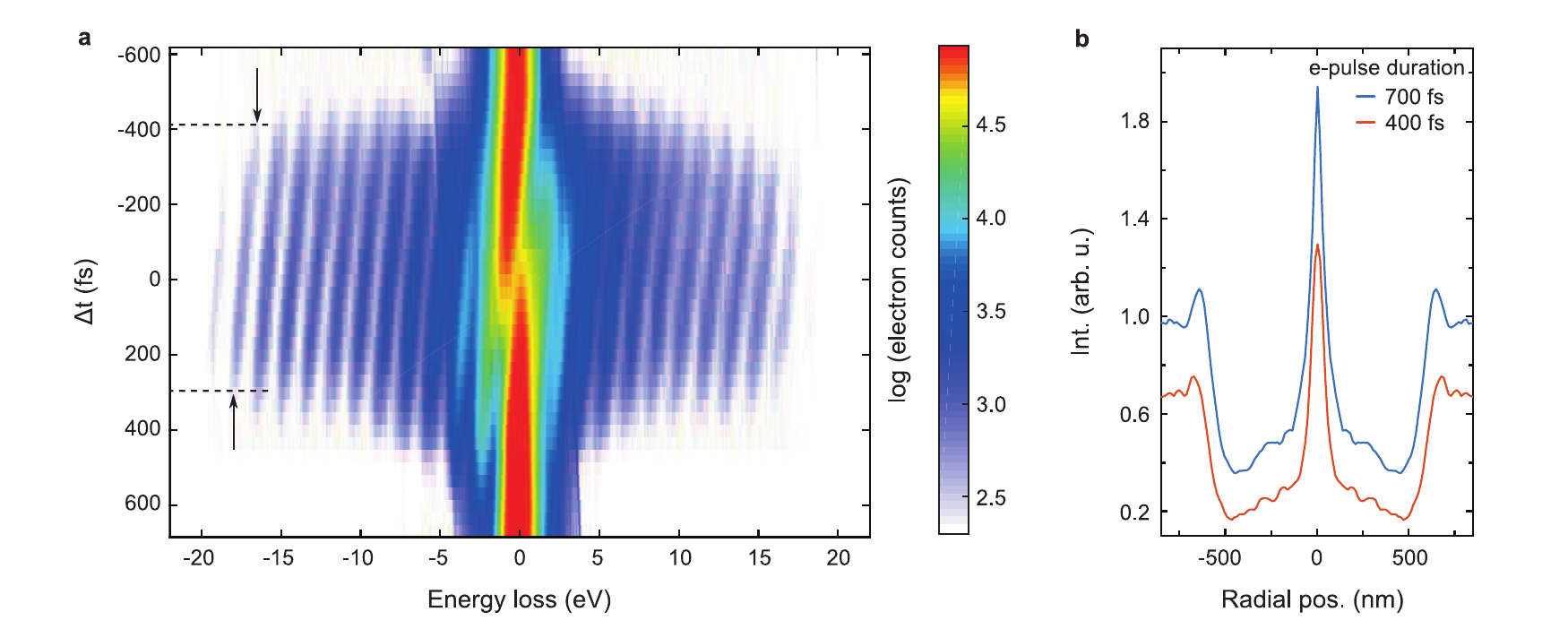}
	\caption{Experimental determination of electron pulse duration and Lorentz imaging with higher temporal resolution. (a) Electron energy-loss spectra as a function of delay time ${\Delta t}$ (close to ${\Delta t = 0}$) for the temporal characterization of employed electron pulses. Arrows: extracted electron pulse duration. (b) Profiles of Lorentz micrographs of the magnetic disc obtained using 400-fs and 700-fs electron pulses at a fixed delay time (\mbox{-5 ps}), respectively.}
	\label{supplfig:duration}
\end{center}
\end{figure}

For synchronizing laser-pump and electron-probe pulses, and for characterizing the electron pulses duration, we make use of electron-optical cross-correlation in laser-induced near-fields \cite{barwick2009, feist2015}. At delay times for which optical-pump pulse and electron-probe pulses overlap at the sample surface, the electron energy spectra exhibit sidebands separated by the photon energy, as shown in Fig.~\ref{supplfig:duration}a. The 700-fs duration of the electron probe pulses can be extracted from the temporal width of the highest-order photon-sidebands \cite{barwick2009, feist2017}.

We note that a light-induced change in electron energy width of about 20 eV during the temporal laser-electron overlap does not  lead to a significant variation in Lorentz contrast. However, the out-of-focus imaging conditions adopted in Lorentz  microscopy are particularly sensitive to modifications in the transverse electron momentum (i.e. perpendicular to the propagation direction), which can be modulated by the interaction of electrons with optical near-fields. Considering the pump photon momenta, electron deflections on the order of 10~$\mu$rad are expected, yielding streak-like 50-nm contrast features in Lorentz micrograph at \mbox{5-mm} defocus (Feist et al., in preparation). In the experiments reported here, such side effects are largely absent, since the laser pump pulse is much shorter than the electron pulse.

In the main text, we demonstrated the use of 700-fs electron pulses for imaging magnetization dynamics with 95-nm spatial resolution, and the possibility to further reduce the spatial resolution to 55~nm. Finally, we discuss the possibilities for improving the temporal resolution by using shorter electron pulses. Figure~\ref{supplfig:duration}b shows image profiles of the magnetic vortex for a fixed delay time of \mbox{-5 ps} and a pump fluence of \mbox{3.7 mJ/cm$^2$}, using the same imaging conditions as for the time-resolved experiments, when probing the sample with \mbox{700-fs} and \mbox{400-fs} electron pulses. Each image contains the same total electron dose of ${1.5\times10^7}$ electrons, consisting of a sum of 6 individual exposures of 10 s each for the longer electron pulses, and of 12 individual exposures for the shorter electron pulses, respectively. The spatial resolution of the magnetic contrast feature is not visibly affected by the lower electron number per pulse, so that time-resolved Lorentz microscopy experiments with higher temporal resolution are clearly feasible.

\addcontentsline{toc}{section}{References}
\bibliographystyle{ieeetr}
\bibliography{lorentzUTEM}

\end{document}